\documentclass[conference]{IEEEtran}
\IEEEoverridecommandlockouts
\usepackage{tex/packages}
\usepackage{flushend}

\addto\captionsenglish{\renewcommand{\figurename}{Fig.}}

\begin{document}
\begin{acronym}
    \acro{CBADC}{control-bounded analog-to-digital converter}
    \acro{CTSDM}[CT-$\Sigma\Delta$M]{continuous-time sigma-delta modulator}
    \acro{CTSD}{continuous-time sigma-delta}
    \acro{DTSDM}{discrete-time sigma-delta modulator}
    \acro{CRFB}{cascade of resonators with feedback}
    \acro{CRFF}{cascade of resonators with feedforward}

    \acro{ENOB}{effective number of bits}
    \acro{SNR}{signal-to-noise ratio}
    \acro{SNDR}{signal-to-noise-and-distortion ratio}
    \acro{SQNR}{signal-to-quantization-noise ratio}
    \acro{SC}{switched-capacitor}
    \acro{CT}{continuous-time}
    \acro{DT}{discrete-time}
    \acro{PSD}{power spectral density}

    \acro{NTF}{noise transfer function}
    \acro{STF}{signal transfer function}

    \acro{CI}{chain-of-integrators}
    \acro{LF}{leapfrog}
    \acro{DC}{digital control}
    \acro{AS}{analog system}
    \acro{DE}{digital estimator}

    \acro{RZ}{return-to-zero}
    \acro{NRZ}{non-return-to-zero}
    \acro{SCR}{switched-capacitor-resistor}
    \acro{DAC}{digital-to-analog converter}
    \acro{ADC}{analog-to-digital converter}

    \acro{FIR}{finite impulse response}
    \acro{FS}{full scale}
    \acro{OSR}{oversampling ratio}
    \acro{OTA}{operational transconductance amplifier}
    \acro{PLL}{phase-locked loop}
    \acro{VCO}{voltage-controlled oscillator}
    \acro{PM}{phase-modulated}
    \acro{FM}{frequency-modulated}
    
    \acro{GBWP}{gain-bandwidth product}
    \acro{PLL}{phase-locked loop}
    \acro{LMS}{least mean squares}
    \acro{RLS}{recursive least squares}
    \acro{AF}{analog frontend}
    \acro{SDM}[$\Sigma\Delta$M]{sigma-delta modulator}
    \acro{QSDM}[Q-$\Sigma\Delta$M]{quadrature sigma-delta modulator}
    \acro{A/D}{analog-to-digital}
    % \acro{BPSDM}{Band-Pass Sigma-Delta Modulator}
    \acro{BPSDM}[BP-$\Sigma\Delta$M]{band-pass sigma-delta modulator}
    \acro{RF}{radio frequency}
    \acro{CT}{continuous-time}
    \acro{DT}{discrete-time}
    \acro{SDR}{software-define-radio}
    \acro{QCBADC}[Q-CBADC]{quadrature control-bounded analog-to-digital converter}
    \acro{BW}{bandwidth}
\end{acronym}

\title{Quadrature Control-Bounded ADCs}

\author{\IEEEauthorblockN{Hampus Malmberg$^1$, Fredrik Feyling$^2$, and Jose M de la Rosa$^3$}
\IEEEauthorblockA{
\textit{Dept. of Information Technology \& Electrical Engineering, ETH Zürich, Zürich, Switzerland$^1$} \\
\textit{Dept. of Electronic Systems, Norwegian University of Science and Technology, Trondheim, Norway$^2$} \\
\textit{Institute of Microelectronics of Seville, IMSE-CNM (CSIC/University of Seville), Seville, Spain$^3$}
}
}

\IEEEoverridecommandlockouts
\IEEEpubid{\begin{minipage}{\textwidth}\
\copyright~2023 IEEE. 
Personal use of this material is permitted.  
Permission from IEEE must be obtained for all other uses, in any current or future media, including reprinting/republishing this material for advertising or promotional purposes, creating new collective works, for resale or redistribution to servers or lists, or reuse of any copyrighted component of this work in other works
\end{minipage}
}

\maketitle

\begin{abstract}
In this paper, the design flexibility of the control-bounded analog-to-digital converter principle is
demonstrated by considering band-pass analog-to-digital conversion.
We show how a low-pass control-bounded analog-to-digital converter can be translated into a band-pass version where the guaranteed stability, converter bandwidth, and signal-to-noise ratio are preserved while the center frequency for conversion can be positioned freely. The proposed converter is validated with behavioral simulations on several filter orders, center frequencies, and oversampling ratios. 
Finally, robustness against component variations is demonstrated by Monte Carlo simulations.
\end{abstract}


\section{Introduction}

\Acp{BPSDM} allow digitizing non-base-band signals; essentially expediting the role and position of \ac{A/D} conversion %, for converting \ac{RF} signals, 
in the receiving structure of wireless receivers.
A mainly digital wireless receive path is beneficial as digital signal processing offers better technology scaling and programmability towards a
\ac{SDR} platform \cite{saye20, ghae21, jie21}.
Digitizing \ac{RF} signals typically utilize sampling frequencies in the GHz range. Hence, state-of-the-art \ac{BPSDM}s are primarily implemented using \ac{CT} circuits as they offer inherent anti-aliasing filtering and are potentially faster and more power efficient than their \ac{DT} counterparts. However, in the majority of cases, \ac{RF} \acp{BPSDM} have a fixed ratio between the center or \textit{notch} frequency, $f_n$, and the sampling frequency $f_s$ (typically $f_n=f_s/4$).
A fixed $f_s/f_n$ ratio results in two main limitations: firstly, for wireless standards operating around 2.5-5GHz, prohibitive values of $f_s$, in the order of tens of GHz, are typically required. Secondly, a widely programmable \ac{PLL} is required for tuning $f_n$ while keeping the $f_s/f_n$ ratio fixed.
These limitations have prompted the interest in reconfigurable \ac{BPSDM}s with tunable notch
frequency \cite{moli14}. However, reported solutions are limited in practice by the increased (analog) circuit complexity and risk of the potential instability of the loop filter -- compromised by the tuning range of $f_n$\cite{moli14}.
This paper presents an alternative approach to the problem of digitizing \ac{RF} signals using the so-called \ac{CBADC} concept \cite{M:20,MWL:21}. A \ac{QCBADC} is proposed, that offers a highly modular architecture with a tunable $f_n$ and a stability guarantee. In particular, the \ac{QCBADC} follows from extending two low-pass \acp{CBADC} into a single oscillating structure. Conveniently, the \ac{QCBADC}'s \ac{SNR} and \ac{BW} specification follows from its two low-pass \acp{CBADC} building blocks.
Like \acp{QSDM} \cite{ST:05}, the \acp{QCBADC} is a quadrature \ac{ADC} resulting in the same number of integrating stages per signal as its low-pass building block. 

\IEEEpubidadjcol
\section{The Leapfrog Analog Frontend}\label{sec:leapfrog}
% The key building block of the quadrature Leapfrog \ac{ADC} is the Leapfrog analog frontend, whose principle operation is described in this section.

\begin{figure*}[tbp]
    \begin{center}
        \resizebox{\textwidth}{!}{
            \input{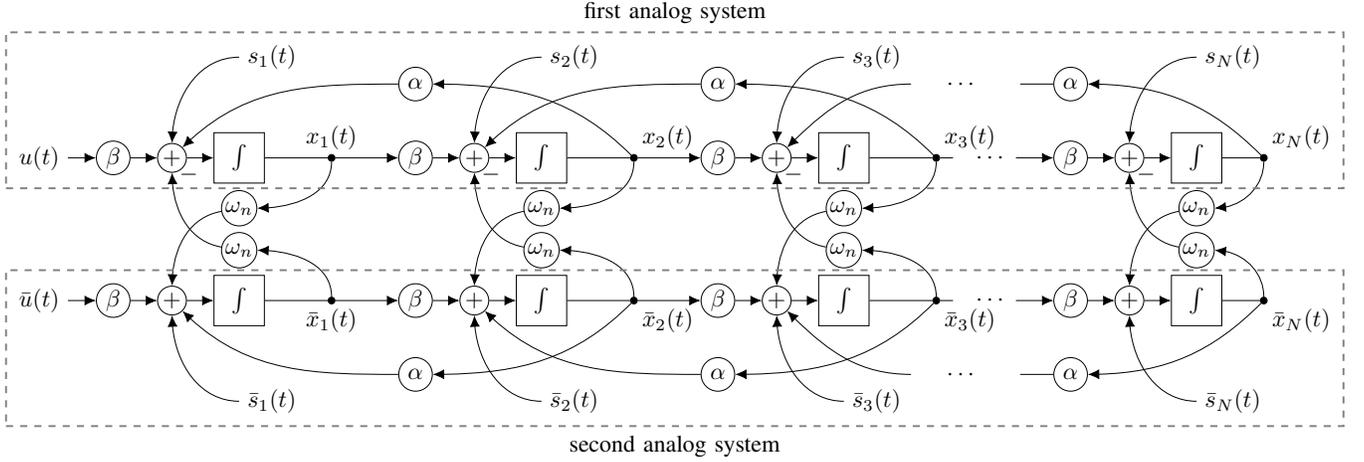}
        }
        \caption{\label{fig:leapfrog-structure} The quadrature Leapfrog analog system which is is the combination of two Leapfrog structures, as in \Sec{sec:leapfrog},
        connected by the $\omega_n$ paths. The system is stabilized via the control signals $s_1(t),\dots,s_N(t),\bar{s}_1(t),\dots,\bar{s}_N(t)$ resulting from $N$ quadrature local digital controls as shown in \Fig{fig:digital_control_zero_order_hold}.
        }
        
    \end{center}
\end{figure*}

The \ac{CBADC} concept builds on the idea that an \ac{AS} stabilized by a \ac{DC} amounts to an implicit \ac{A/D} conversion.
As an example of such an \ac{AS}, consider the system within one of the dashed boxes in \Fig{fig:leapfrog-structure}.
This example shows a low-pass \ac{LF} \ac{AS} \cite{M:20}, parameterized by a forward and feedback gain $\beta$ and $\alpha$.
To turn the \ac{AS} into an \ac{ADC}, it's stabilized by a \ac{DC} through the control signals $s_1(t), \dots, s_N(t)$. A \ac{CBADC}'s overall conversion performance follows from the open-loop gain of the \ac{AS} in combination with a bounded state swing on the state variables $x_\ell(t)$ enforced by the \ac{DC}—particularly, a large \ac{AS} gain in combination with a, \ac{DC} enforced, small state swing result in good conversion performance.

In a subsequent step, the \ac{CBADC}'s final output follows from the control signals as % the discrete-time convolution of $N$ \ac{FIR} filters as
\begin{IEEEeqnarray}{rCl}
    \hat{u}[k]  & \eqdef & \sum_{\ell = 1}^N (h_\ell \ast s_\ell)[k]. \label{eq:estimate}
    %  & = & \sum_{\ell = 0}^N \vct{s}_\ell(k)^\T \vct{h}_\ell
\end{IEEEeqnarray}
where the \ac{FIR} filter coefficients $h_1[.], \dots, h_\ell[.]$, in \Eq{eq:estimate}, depend on the parametrization of the \ac{AF}, i.e., the combined \ac{AS} and \ac{DC}. Furthermore, the discrete time control signals in \Eq{eq:estimate} are
related to their continuous time counterparts as 
\begin{IEEEeqnarray}{rCl}
    s_\ell(t) & = & \sum_k s_\ell[k] \theta_\ell(t - kT)
\end{IEEEeqnarray}
where the \ac{DC} updates the control signals with a clock frequency of $f_s\eqdef1/T$ and $\theta(t)$ is the corresponding \ac{DAC}  waveform.
The FIR filters $h_1[.],\dots,h_N[.]$ in \Eq{eq:estimate} can be calculated analytically as in \cite{MWL:21} or via calibration as in \cite{MMBFL:22}. Beware that component variation in the \ac{AS} or \ac{DC} will introduce a significant estimation error if not accounted for in $h_1[.],\dots,h_N[.]$, cf. \cite{FMWLY:2022} and \cite{MMBFL:22}. 

% As will become apparent in \Sec{sec:quadrature}, the \ac{LF} \ac{QCBADC} 
The \ac{AS} is conveniently described using state-space equations;
an $N$th order \ac{LF} \ac{AS}, follows from the differential equations
\begin{IEEEeqnarray}{rCl}
    \dot{\vct{x}}(t) & = & \mat{A}_{\text{LF}} \vct{x}(t) + \mat{B}_{\text{LF}} u(t) + \vct{s}(t) \label{eq:state_equations_leapfrog} %\\
    % \tilde{\vct{s}}(t) & = & \tilde{\mat{\Gamma}}^\T \vct{x}(t) + \tilde{\mat{B}}^\T u(t) + \tilde{\mat{A}}^\T \vct{s}(t) \label{eq:control_observation}
\end{IEEEeqnarray}
where
% $\vct{x}(t) \eqdef \begin{pmatrix}x_1(t), \dots, x_N(t)\end{pmatrix}^\T \in \R^N$ and $\vct{s}(t) \eqdef \begin{pmatrix}s_0(t), \dots, s_N(t)\end{pmatrix}^\T \in \{\pm1\}^N$ denote the %$N$ dimensional 
% the analog state vector 
% $\vct{x}(t) \eqdef \begin{pmatrix}x_1(t), \dots, x_N(t)\end{pmatrix}^\T$,
% the control signal contribution vector 
% $\vct{s}(t) \eqdef \begin{pmatrix}s_1(t), \dots, s_N(t)\end{pmatrix}^\T$, %denote the $N$ dimensional,
% analog state-vector and control signal contribution vector, respectively,
\begin{IEEEeqnarray}{rCl}
    \vct{x}(t) &\eqdef& \begin{pmatrix}x_1(t), \dots, x_N(t)\end{pmatrix}^\T, \\
    \vct{s}(t) &\eqdef& \begin{pmatrix}s_1(t), \dots, s_N(t)\end{pmatrix}^\T, \\
    \mat{A}_{\text{LF}} & \eqdef &
    \begin{pmatrix}
        0 & \alpha \\
        \beta & 0 & \ddots \\ 
         & \ddots & \ddots & \alpha \\
         &  & \beta & 0  \\
    \end{pmatrix} \in \R^{N \times N},
\end{IEEEeqnarray}
and $\mat{B}_{\text{LF}} \eqdef \begin{pmatrix}\beta, 0,\dots, 0\end{pmatrix}^\T\in \R^N$.
% For a low-pass Leapfrog \ac{ADC} with a bandwidth $\wb$, it was shown in \cite{M:20} that the \ac{SNR} may be approximated as
% \begin{IEEEeqnarray}{rCl}
%     \label{eq:snr_expr}
%     \text{SNR} \propto \left(P_{x_N} \int_0^{\wb} |G(\omega)|^{-2}d\omega \right)^{-1},
% \end{IEEEeqnarray}
% where $P_{x_N}$ is the total power of $x_N(t)$. 
% As shown in \cite{M:20}, the \ac{SNR} of an $N$'th order Leapfrog \ac{ADC} relates to the parameter $\beta$ as $\text{SNR} \propto \beta^{2N}$. Furthermore, the system has a guaranteed stability and a signal bandwidth of $\wb$ [rad/s] if $\kappa, \alpha$ and $f_s$ are 
% In contrast to \acp{CTSDM} which are typically designed 

The \ac{LF}'s homogeneous and modular structure enables design equations where %, based on closed form design equations for the parameters $\beta, \alpha$ and $\kappa$ 
for a given $N$, $\wb$, and \ac{OSR},% i.e., $\text{\ac{OSR}} \eqdef \frac{\pi f_s}{\omega_{\mathcal{B}}}$,
\begin{IEEEeqnarray}{rCl}%
\text{SNR} & \propto & \left(\text{OSR}\right)^{2N} \labell{eq:SNR_scaling} 
\end{IEEEeqnarray}
% where the proportionality relation is determined by a single empirical constant \cite{M:20}. 
and stability %and the converters's bandwidth 
follow from the relations
\begin{IEEEeqnarray}{rCl}
    |\beta| & = & \frac{\wb \cdot \text{OSR}}{2 \pi} \labell{eq:stability} \\
    \kappa & = & -\beta = \frac{\wb^2}{4\alpha} \labell{eq:leapfrog_relation}
\end{IEEEeqnarray}
where 
% the \ac{OSR} is defined as $\text{OSR} \eqdef \frac{\pi f_s}{\omega_{\mathcal{B}}}$ and 
$\wb$ [rad/s] is the targeted angular signal bandwidth.
\cite{FMWLY:2022} showed that the nominal performance of a low-pass \ac{LF} \ac{CBADC} is similar to that of a heuristically optimized \ac{CTSDM} with the same loop filter order $N$, fixed \ac{OSR}, and the same number of quantization levels. 

\section{Quadrature Analog Frontends}
\label{sec:quadrature}
Two low-pass \acp{CBADC}, as in \Sec{sec:leapfrog}, can be turned into a quadrature \ac{CBADC} by two modifications:
firstly, interconnecting the two \acp{AS} such that they oscillate at the desired notch frequency $f_n$, further described in \Sec{sec:qas}.
Secondly, the resulting \ac{AS} can be stabilized by a local quadrature \ac{DC}, covered in \Sec{sec:lqdc}. % as given in \Fig{fig:digital_control_zero_order_hold}. 
This general principle applies to any low-pass \ac{CBADC} \ac{AF}. 
In the interest of space, only the transformation of the low-pass \ac{LF} from \Sec{sec:leapfrog} will be covered in this paper.

\subsection{Quadrature Analog System}\label{sec:qas}
For a quadrature \ac{AS} to oscillate at a desired angular frequency $\omega_n = 2 \pi f_n$,
two identical $N$th order \acp{AS} are stacked in parallel and interconnected as
\begin{IEEEeqnarray}{rCl}%
    \begin{pmatrix}\dot{\vct{x}}(t) \\ \dot{\bar{\vct{x}}}(t)\end{pmatrix} & = & \mat{A} \begin{pmatrix}\vct{x}(t) \\ \bar{\vct{x}}(t)\end{pmatrix} + \mat{B}\begin{pmatrix}u(t) \\ \bar{u}(t) \end{pmatrix} + \begin{pmatrix}\vct{s}(t) \\ \bar{\vct{s}}(t)\end{pmatrix}  \labell{eq:state_space_equation} \\
    \mat{A} & \eqdef & \begin{pmatrix} \mat{A}_{\text{LP}} & -\omega_n \mat{I}_N \\ \omega_n \mat{I}_N &  \mat{A}_{\text{LP}} \end{pmatrix} \in \R^{2 N \times 2 N} \\
    \mat{B} & \eqdef & \begin{pmatrix} \mat{B}_{\text{LP}} & \mat{0}_N \\ \mat{0}_N & \mat{B}_{\text{LP}} \end{pmatrix} \in \R^{2 N \times 2}  \label{eq:ssm_quad_3}
\end{IEEEeqnarray}
where $\mat{A}_{\text{LP}}$ and $\mat{B}_{\text{LP}}$ refers to the system description of a low-pass \ac{AS}, e.g., the $\mat{A}_{\text{LF}}$ and $\mat{B}_{\text{LF}}$ from \Sec{sec:leapfrog}, and $\vct{x}(t)$, $\bar{\vct{x}}(t)$, and $\vct{s}(t)$, $\bar{\vct{s}}(t)$ are the in-phase and quadrature part of the state vector and control signal vector, respectively.
\Fig{fig:leapfrog-structure}, shows the low-pass \ac{LF}, from \Sec{sec:leapfrog}, transformed into its quadrature version.

\subsection{Local Quadrature Digital Control}\label{sec:lqdc}
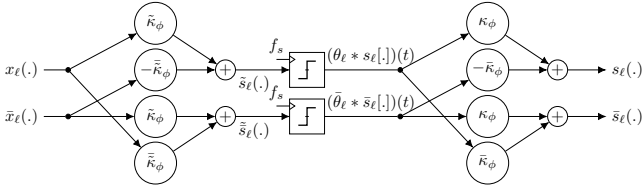
\begin{figure}[tbp]
    \begin{center}
        \resizebox{\columnwidth}{!}{
            \begin{tikzpicture}[node distance=1.25cm,AnalogMultiplier/.style={AnalogSum, font={}, minimum size=0.9cm}]

    %%%%%%%%%%%%%%%%%%%%%%%%%%%%%%%%%%%%%%%%%%%%%%%%%%%%%%%%%%%%%%%
    % First Leap Frog
    %%%%%%%%%%%%%%%%%%%%%%%%%%%%%%%%%%%%%%%%%%%%%%%%%%%%%%%%%%%%%%%

    % First Node
    \node (x_tilde_I) {$x_\ell(.)$};
    \node (x_tilde_Q) at ($(x_tilde_I) + (0, -1)$) {$\bar{x}_\ell(.)$};

    \node[AnalogBranch] (b1) at ($(x_tilde_I) + (1,0)$) {} edge[] (x_tilde_I);
    \node[AnalogBranch] (b2) at ($(x_tilde_Q) + (1,0)$) {} edge[] (x_tilde_Q);
    
    \node[AnalogMultiplier] (m1) at ($(b1) + (1.875,1)$) {$\tilde{\kappa}_\phi$};
    \node[AnalogMultiplier] (m2) at ($(b1) + (1.875,0)$) {$-\bar{\tilde{\kappa}}_\phi$};
    \node[AnalogMultiplier] (m3) at ($(b1) + (1.875,-1)$) {$\tilde{\kappa}_\phi$};
    \node[AnalogMultiplier] (m4) at ($(b1) + (1.875,-2)$) {$\bar{\tilde{\kappa}}_\phi$};

    % \node (kappa_tilde_1) at ($(x_tilde_I) + (0, 1)$) {$\tilde{\kappa}_\phi$};
    % \node[AnalogBranch] (b3) at ($(kappa_tilde_1) + (1,0)$) {} edge[] (kappa_tilde_1);
    % \draw[Arrow] (b3) -- (m1);
    % \draw[Arrow] (b3) -- (m2);

    % \node (kappa_tilde_2) at ($(x_tilde_Q) + (0, -1)$) {$\bar{\tilde{\kappa}}_\phi$};
    % \node[AnalogBranch] (b4) at ($(kappa_tilde_2) + (1,0)$) {} edge[] (kappa_tilde_2);
    % \node[AnalogMultiplier] (neg) at ($(b4) + (0.875,0)$) {-1};
    % \draw[Arrow] (b4) -- (neg); 
    % \draw[Arrow] (neg) -- (m4);
    % \draw[Arrow] (b4) -- (m3);

    \node[AnalogSum] (sum1) at ($(m2) + (1.5, 0)$) {};
    \node[AnalogSum] (sum2) at ($(m3) + (1.5, 0)$) {};

    \draw[Arrow] (m1) -- (sum1);
    \draw[Arrow] (m2) -- (sum1);
    \draw[Arrow] (m3) -- (sum2);
    \draw[Arrow] (m4) -- (sum2);

    % \draw[Arrow] (b3) -- (m1);
    % \draw[Arrow] (b3) -- (m3);
    % \draw[Arrow] (b4) -- (m2);
    % \draw[Arrow] (b4) -- (neg);
    % \draw[Arrow] (neg)-- (m4);

    \draw[Arrow] (b1) -- (m1);
    \draw[Arrow] (b1) -- (m4);
    \draw[Arrow] (b2) -- (m2);
    \draw[Arrow] (b2) -- (m3);
    
    \draw pic[rotate=0, yscale=1] (quantizer1) at ($(sum1) + (1.75, 0)$) {quantizer};
    
    \draw[Arrow] (sum1) -- (quantizer1-in);
    
    \draw pic[rotate=0, yscale=1] (quantizer2) at ($(sum2) + (1.75, 0)$) {quantizer};
    
    \draw[Arrow] (sum2) -- (quantizer2-in);
    % \draw[Arrow] (quantizer2-out) -- (s_Q);

    \node[AnalogBranch] (b6) at ($(quantizer1) + (2,0)$) {} edge[] (quantizer1-out);
    \node[AnalogBranch] (b7) at ($(quantizer2) + (2,0)$) {} edge[] (quantizer2-out);

    % \node (kappa_1) at ($(quantizer1) + (0, 1)$) {$\kappa_\phi$};
    % \node[AnalogBranch] (b5) at ($(kappa_1) + (2,0)$) {} edge[] (kappa_1);

    % \node (kappa_2) at ($(quantizer2) + (0, -1)$) {$\bar{\kappa}_\phi$};
    % \node[AnalogBranch] (b8) at ($(kappa_2) + (2,0)$) {} edge[] (kappa_2);

    \node[AnalogMultiplier] (m5) at ($(b6) + (1.875,1)$) {$\kappa_\phi$};
    \node[AnalogMultiplier] (m6) at ($(b6) + (1.875,0)$) {$-\bar{\kappa}_\phi$};
    \node[AnalogMultiplier] (m7) at ($(b7) + (1.875,0)$) {$\kappa_\phi$};
    \node[AnalogMultiplier] (m8) at ($(b7) + (1.875,-1)$) {$\bar{\kappa}_\phi$};

    % \node[AnalogMultiplier] (neg) at ($(b8) + (0.875,0)$) {-1};

    \node[AnalogSum] (sum3) at ($(m6) + (1.5, 0)$) {};
    \node[AnalogSum] (sum4) at ($(m7) + (1.5, 0)$) {};

    \draw[Arrow] (m5) -- (sum3);
    \draw[Arrow] (m6) -- (sum3);
    \draw[Arrow] (m7) -- (sum4);
    \draw[Arrow] (m8) -- (sum4);

    % \draw[Arrow] (b5) -- (m5);
    % \draw[Arrow] (b5) -- (m7);
    % \draw[Arrow] (b8) -- (m6);
    % \draw[Arrow] (b8) -- (neg);
    % \draw[Arrow] (neg)-- (m8);

    \draw[Arrow] (b6) -- (m5);
    \draw[Arrow] (b6) -- (m8);
    \draw[Arrow] (b7) -- (m6);
    \draw[Arrow] (b7) -- (m7);

    \node (s_I) at ($(sum3) + (1.5,0)$) {$s_\ell(.)$} edge[BackArrow] (sum3);
    \node (s_Q) at ($(sum4) + (1.5,0)$) {$\bar{s}_\ell(.)$} edge[BackArrow] (sum4);

    \node[anchor=north] at ($(sum1) + (0.615, 0)$) {$\tilde{s}_\ell(.)$};
    \node[anchor=north] at ($(sum2) + (0.615, 0)$) {$\tilde{\bar{s}}_\ell(.)$};

    \node[anchor=south] at ($(b6) - (0.625, 0)$) {$(\theta_\ell \ast s_\ell[.])(t)$};
    \node[anchor=south] at ($(b7) - (0.625, 0)$) {$(\bar{\theta}_\ell \ast \bar{s}_\ell[.])(t)$};

\end{tikzpicture}
        }
        \caption{\label{fig:digital_control_zero_order_hold} The $\ell$th local quadrature \ac{DC}, connecting $\vct{x}_\ell(t) = (x_\ell(t), (\bar{x}_\ell(t))$ with $\vct{s}_\ell = (s_\ell(.),\bar{s}_\ell(.))$ in \Fig{fig:leapfrog-structure}. 
        The output of the two comparators are considered continuous-time quantities $((\theta_\ell \ast s_\ell[.])(t), (\bar{\theta}_\ell \ast \bar{s}_\ell[.])(t))$, where $(\theta_\ell(.), \bar{\theta}_\ell(.))$ are the comparators' impulse responses and $(s[.], \bar{s}[.])$ are the discrete-time control decisions used in \Eq{eq:estimate}.}
    \end{center}
\end{figure}
The quadrature \ac{AS}, from \Sec{sec:qas}, can be stabilized by $N$ local quadrature \acp{DC} as shown in \Fig{fig:digital_control_zero_order_hold}. 
Here each quadrature state pair $\vct{x}_\ell(t) \eqdef \begin{pmatrix}x_\ell(t), \bar{x}_\ell(t)\end{pmatrix}^\T$
are turned into a control observation
\begin{IEEEeqnarray}{rCl}%
    \tilde{\vct{s}}_\ell(t) & \eqdef & \begin{pmatrix}\tilde{s}_\ell(t) \\  \bar{\tilde{s}}_\ell(t)\end{pmatrix} = \begin{pmatrix}\tilde{\kappa}_\phi & - \bar{\tilde{\kappa}}_\phi \\ \bar{\tilde{\kappa}}_\phi  & \tilde{\kappa}_\phi\end{pmatrix} \vct{x}_\ell(t) \label{eq:control_observation}
\end{IEEEeqnarray}
which is then sampled and quantized into the quadrature discrete-time control signal pair $\vct{s}_\ell[k] \eqdef \begin{pmatrix}s_\ell[k], \bar{s}_\ell[k]\end{pmatrix}^\T$. For a non-return to zero DAC the $\ell$th quadrature control contribution pair follows as
\begin{IEEEeqnarray}{rCl}%
    \vct{s}_\ell(t) & \eqdef & \begin{pmatrix}s_\ell(t) \\ \bar{s}_\ell(t)\end{pmatrix} = \begin{pmatrix}\kappa_\phi & -\bar{\kappa}_\phi \\ \bar{\kappa}_\phi & \kappa_\phi\end{pmatrix} \vct{s}_\ell[k] \label{eq:control_signal}
\end{IEEEeqnarray}
where $t \in ((k-1)T + \tau_{\text{DC}}, kT + \tau_{\text{DC}}]$ and $\tau_{\text{DC}}>0$ is the time delay associated with the quantizer.

To determine the required values of $\kappa_\phi$, $\bar{\kappa}_\phi$, $\tilde{\kappa}_\phi$, and $\tilde{\bar{\kappa}}_\phi$ %, cf. \Eq{eq:control_observation} and \Eq{eq:control_signal}, 
that stabilize the quadrature \ac{AS}, the derivation of the local low-pass \ac{DC}, and its stability guarantee, in \cite{M:20} is extended into the quadrature case.
Specifically, $(\kappa_\phi,\bar{\kappa}_\phi,\tilde{\kappa}_\phi,\tilde{\bar{\kappa}}_\phi, T)$ are chosen such that, at the end of a control-period $T$, each quadrature state pair $\|\vct{x}_\ell(T)\|_2$ is bounded by a positive constant $\epsilon$ for any quadrature input pair $\|\vct{x}_{\ell-1}(t)\|_2 < \epsilon$, and initial state $\|\vct{x}_\ell(0)\|_2 < \epsilon$, where $t\in[0, T)$. It follows that if
such a parametrization exists, the system as a whole will be inherently stable by a recursive argument given the first quadrature input pair $\|\vct{x}_0(t)\|_2 \eqdef \|\begin{pmatrix}u(t), \bar{u}(t)\end{pmatrix}\|_2 < \epsilon$.

These general conditions can be reduced to design equations for the involved parameter values. The steps involved include analytical solutions to differential equations systems and general matrix properties. Given the space limitation, only the resulting expressions will be presented. Simulation results will be shown in \Sec{sec:simulation} to illustrate and validate the relevance of these mathematical expressions. %% JMR

\subsubsection{Matched Signal Strengths}
Matching the largest control and input signal contribution reduces to the condition
\begin{IEEEeqnarray}{rCl}%
    \sqrt{\kappa_\phi^2 + \bar{\kappa}_\phi^2} & \overset{!}{=} & \frac{\beta T \omega_n}{2 \sin\left(\frac{\omega_n T}{2}\right)}. \label{eq:control_gain}
\end{IEEEeqnarray}

\subsubsection{Self Stability}
Anticipating and aligning the rotating state trajectories at the end of a control period results in the two conditions
\begin{IEEEeqnarray}{rCl}%
    \sqrt{\tilde{\kappa}_\phi^2 + \bar{\tilde{\kappa}}_\phi^2} & \overset{!}{=} & \frac{\omega_n}{2 \sqrt{\kappa_\phi^2 + \bar{\kappa}_\phi^2} \sin\left(\frac{\omega_n T}{2}\right)} \label{eq:kappa_tilde_scaling} \\
    \arctan\left(\frac{\bar{\tilde{\kappa}}_\phi}{\tilde{\kappa}_\phi}\right) & \overset{!}{=} & \omega_n \left(\frac{T}{2} + \tau_{\text{DC}}\right) - \phi_\kappa + \pi, \label{eq:kappa_tilde_angle}
\end{IEEEeqnarray}
where $\phi_\kappa = \arctan\left(\bar{\kappa}_\phi / \kappa_\phi\right)$.
By combining \Eq{eq:control_gain}, \Eq{eq:kappa_tilde_scaling}, and \Eq{eq:kappa_tilde_angle} follows
\begin{IEEEeqnarray}{rCl}%
    \kappa_\phi & = &  \frac{\beta T \omega_n}{2 \sin\left(\frac{\omega_n T}{2}\right)}\cos\left(\phi_\kappa\right) \label{eq:kappa_phi}\\
    \bar{\kappa}_\phi & = &  \frac{\beta T \omega_n}{2 \sin\left(\frac{\omega_n T}{2}\right)} \sin\left(\phi_\kappa\right) \\
    \tilde{\kappa}_\phi & = & -\frac{1}{\beta T}\cos\left(\omega_n \left(\frac{T}{2} + \tau_{\text{DC}}\right) - \phi_\kappa\right)\\
    \bar{\tilde{\kappa}}_\phi & = & -\frac{1}{\beta T}\sin\left(\omega_n \left(\frac{T}{2} + \tau_{\text{DC}}\right) - \phi_\kappa\right) \label{eq:bar_tilde_kappa_phi}
\end{IEEEeqnarray}
where $\phi_\kappa\in[0, 2\pi)$ is a free parameter that may be chosen to ensure practical values.

\subsubsection{Worst-Case Superposition}
\begin{figure}
    \centering
    \begin{tikzpicture}
        \pgfplotsset{
            grid=both, 
            % minor grid style = {densely dotted}, 
            % major grid style = {densely dotted}
        }
        \pgfplotsset{ylabel near ticks, xlabel near ticks}
        \pgfplotsset{ 
            legend cell align={left},
            legend style={
                at={(0.985,0.98)},
                anchor=north east,
                fill opacity=0.85,
                draw opacity=1,
                text opacity=1,
                nodes={scale=0.8, transform shape}
            }
        }
        \begin{axis}[
            xlabel=$\omega_n T / (2 \pi)$,
            ylabel=Coeff. value,
            xmin=0,
            xmax=0.5,
            legend pos=north west,
            width=0.95\columnwidth,
            % height=\columnwidth/1.618,
            height=\columnwidth/2.6,
            xtick = {0, 0.0625, 0.125, 0.1875, 0.25, 0.3125, 0.375, 0.4375, 0.5},
            xticklabels = {$0$, ,$1/8$, ,$1/4$, ,$3/8$, , $1/2$},
            ]
            \addplot[color=black,thick] table[col sep=comma, x=fpT, y=kappa] {./figures/kappa_scaling_v2.csv};
            \addplot[color=black,thick, dashed] table[col sep=comma, x=fpT, y=bar_kappa] {./figures/kappa_scaling_v2.csv};
            \addplot[color=black,thick,dotted] table[col sep=comma, x=fpT, y=tilde_kappa] {./figures/kappa_scaling_v2.csv};
            \addplot[color=black,thick, dash dot] table[col sep=comma, x=fpT, y=bar_tilde_kappa] {./figures/kappa_scaling_v2.csv};
            
            % \addplot[color=blue,thick] table[col sep=comma, x=fpT, y=kappa] {./figures/kappa_scaling_v3.csv};
            % \addplot[color=blue,thick, dashed] table[col sep=comma, x=fpT, y=bar_kappa] {./figures/kappa_scaling_v3.csv};
            % \addplot[color=blue,thick,dotted] table[col sep=comma, x=fpT, y=tilde_kappa] {./figures/kappa_scaling_v3.csv};
            % \addplot[color=blue,thick, dash dot] table[col sep=comma, x=fpT, y=bar_tilde_kappa] {./figures/kappa_scaling_v3.csv};
            
            % \addplot[color=red,thick] table[col sep=comma, x=fpT, y=kappa] {./figures/kappa_scaling_v4.csv};
            % \addplot[color=red,thick, dashed] table[col sep=comma, x=fpT, y=bar_kappa] {./figures/kappa_scaling_v4.csv};
            % \addplot[color=red,thick,dotted] table[col sep=comma, x=fpT, y=tilde_kappa] {./figures/kappa_scaling_v4.csv};
            % \addplot[color=red,thick, dash dot] table[col sep=comma, x=fpT, y=bar_tilde_kappa] {./figures/kappa_scaling_v4.csv};
            \addlegendentry{$\kappa_\phi$}
            \addlegendentry{$\bar{\kappa}_\phi$}
            \addlegendentry{$\tilde{\kappa}_\phi$}
            \addlegendentry{$\bar{\tilde{\kappa}}_\phi$}
        \end{axis}
\end{tikzpicture}
    \caption{\label{fig:kappa_figure} The coefficients in \Eq{eq:kappa_phi}-\Eq{eq:bar_tilde_kappa_phi} as a function of $\omega_n T$ where $2 \beta T = 1$, $\phi_\kappa = 0$, and $\tau_{DC} = 0$.}
\end{figure}
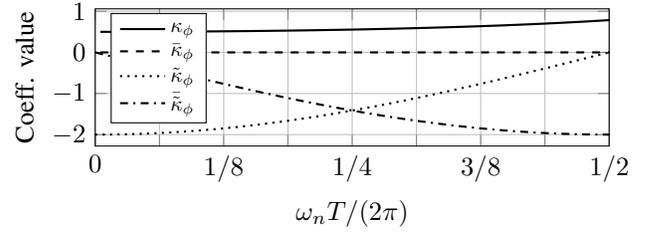

Similarly to the chain-of-integrators case \cite{M:20}, 
\begin{IEEEeqnarray}{rCl}%
    2 \beta T & \leq & 1 \label{eq:superposition}
\end{IEEEeqnarray}
will ensure that the state vector is bounded for a worst-case input and control signal superposition.
\Fig{fig:kappa_figure} demonstrates how $\kappa_\phi$, $\bar{\kappa}_\phi$, $ \tilde{\kappa}_\phi$, and $\tilde{\bar{\kappa}}_\phi$ depends on $\omega_n T$ for $2 \beta T = 1$, $\phi_\kappa = 0$, and $\tau_{DC} = 0$.

In summary, the stability of a quadrature \ac{CBADC} analog frontend, as in \Eq{eq:state_space_equation}-\Eq{eq:ssm_quad_3}, can be ensured 
for $(\kappa_\phi,\bar{\kappa}_\phi,\tilde{\kappa}_\phi,\tilde{\bar{\kappa}}_\phi, T)$ as in \Eq{eq:kappa_phi}-\Eq{eq:superposition}.

\begin{figure}[h]%[tbp]
    \begin{center}
        \resizebox{0.85\columnwidth}{!}{
            % \vspace{-1cm}
\begin{circuitikz}[european voltages]
% Op-amp
\node[op amp, yscale=-1] (amp_I) at (0,0) {\scalebox{1}[-1]{$A$}};
\node[rground] at (amp_I.+) {};
\node[inner sep=0] (vgnd_I) at ($(amp_I.-) + (-0.5,0.5)$) {};
\draw (vgnd_I) to[short, *-] (amp_I.-);
\draw (vgnd_I) to[short, *-] ++(0,-1.25) to[C, l_=$C$] ++ (2.875,0) -| (amp_I.out);

% Quantizer
\node[op amp, yscale=1, rotate=180] (quantizer_I) at ($(amp_I) + (0, 2)$) {};
\draw ($(quantizer_I) + (0.5, -0.25)$) -- ++(-0.25,0) -- ++(0, 0.5) -- ++(-0.25, 0);
\draw ($(quantizer_I) + (0.1875, 0)$) -- ++(0.125,0);
\node (fs) at ($(quantizer_I) + (1.75, 0.675)$) {$f_s$};
\draw[Arrow] (fs) -- ++(-1.1, 0);
\node[rground] at (quantizer_I.+) {};

\draw (amp_I.out) to[R, *-*,l_=$-R_{\bar{\tilde{\kappa}}_I}$] (quantizer_I.-);

% % Alpha Feedback
% \draw (vgnd1) -- ++(-0.5,0.5) to[R, l=$R_{\alpha_1}$] ++(0,1.75) node[inner sep=0] (alpha1) {};

% Op-amp
\node[op amp] (amp_Q) at (0,-5) {$A$};
\node[rground] at (amp_Q.+) {};
\node[inner sep=0] (vgnd_Q) at ($(amp_Q.-) + (-0.5,-0.5)$) {};
\draw (vgnd_Q) to[short, *-] (amp_Q.-);
\draw (vgnd_Q) to[short, *-] ++(0,1.25) to[C, l=$C$] ++ (2.875,0) -| (amp_Q.out);

% Quantizer
\node[op amp, yscale=-1, rotate=180] (quantizer_Q) at ($(amp_Q) + (0, -2)$) {};
\draw ($(quantizer_Q) + (0.5, 0.25)$) -- ++(-0.25,0) -- ++(0, -0.5) -- ++(-0.25, 0);
\draw ($(quantizer_Q) + (0.1875, 0)$) -- ++(0.125,0);
\node (fs) at ($(quantizer_Q) + (1.75, -0.675)$) {$f_s$};
\draw[Arrow] (fs) -- ++(-1.1, 0);
\node[rground] at (quantizer_Q.+) {};

\draw (amp_Q.out) to[R, *-*, l=$R_{\bar{\tilde{\kappa}}_Q}$] (quantizer_Q.-);

\draw (quantizer_I.out) -- ++ (-0.5,0) coordinate (s_I) to[R, l_=$R_{\kappa}$, *-] (vgnd_I);
\draw (quantizer_Q.out) -- ++ (-0.5,0) coordinate (s_Q) to[R, l=$R_{\kappa}$,*-] (vgnd_Q);

\draw (s_I) -- ++(-1,0) -- ++(0,0) to[R, l_=$R_{\bar{\kappa}}$] ++(0,-2) -- ++(0,-2) to[short, -*] ++(0.625,-0.75);
\draw (s_Q) -- ++(-1,0) -- ++(0,0) to[R, l=$-R_{\bar{\kappa}}$] ++(0,2) -- ++(0,2) to[short,-*] ++(0.625,0.75);

\draw (amp_Q.out) -- ++(0.5,0) coordinate (out_Q) to[R, l_=$-R_{\omega_p}$] ++(0,2.25) -- ++(-0.75,0.5) -- ++(-3,0) -- ++(0,1.75) -- (vgnd_I);
\draw (amp_I.out) -- ++(0.5,0) coordinate (out_I) to[R, l=$R_{\omega_p}$] ++(0,-2.25) -- ++(-0.75,-0.5) -- ++(-3,0) -- ++(0,-1.75) -- (vgnd_Q);

\draw (quantizer_I.-) -- ++(2.25,0) to[R, l=$R_{{\tilde{\kappa}}_I}$] ++(0, -1.5) -- ++(0, -2.25) -- ++(-0.5, -0.5) -- ++(0, -2.25) to[short, -*] (out_Q);
\draw (quantizer_Q.-) -- ++(2.25,0) to[R, l_=$R_{{\tilde{\kappa}}_Q}$] ++(0, 1.5) -- ++(0, 2.25) -- ++(-0.5, 0.5) -- ++(0, 2.25) to[short,-*] (out_I); 

\draw ($(vgnd_I) + (-2.75, 0.)$) node[anchor=east] {$v_{x_{\ell-1}}(t)$} to[R,o-,l=$R_{\beta}$] ++(1.75,0) -- (vgnd_I);
\draw ($(vgnd_I) + (-2.75, -0.5)$) node[anchor=east] {$v_{x_{\ell+1}}(t)$} to[R,o-,l_=$R_{\alpha}$] ++(1.75,0) -- (vgnd_I);

\draw ($(vgnd_Q) + (-2.75, 0.)$) node[anchor=east] {$v_{\bar{x}_{\ell-1}}(t)$} to[R,o-,l_=$R_{\beta}$] ++(1.75,0) -- (vgnd_Q);
\draw ($(vgnd_Q) + (-2.75, 0.5)$) node[anchor=east] {$v_{\bar{x}_{\ell+1}}(t)$} to[R,o-,l=$R_{\alpha}$] ++(1.75,0) -- (vgnd_Q);

% \draw ($(vgnd_I) + (-3.25,0)$) node[rground] {} to[V,l=$v_{\bar{x}_{\ell+1}}(t)$] ++(0, 1.25) to[R,l=$R_{\alpha}$] ++(1.75, 0) -- ++(0,-0.75) -- (vgnd_I);

% \draw ($(vgnd_Q) + (-3.75,-1.25)$) node[rground] {} to[V,l=$v_{\bar{x}_{\ell-1}}(t)$] ++(0, 1.25) to[R,l=$R_{\beta}$] ++(1.75,0) -- (vgnd_Q);
% \draw ($(vgnd_Q) + (-3.25,-2.5)$) node[rground] {} to[V,l=$v_{\bar{x}_{\ell+1}}(t)$] ++(0, 1.25) to[R,l=$R_{\alpha}$] ++(1.75, 0) -- ++(0,0.75) -- (vgnd_Q);

% Labels
\node[anchor=south] at (s_I) {$v_{(\theta_\ell \ast s_\ell[.])}(t)$};
\node[anchor=north] at (s_Q) {$v_{(\theta_\ell \ast \bar{s}_\ell[.])}(t)$};

\node[anchor=south west] at (quantizer_I.-) {$v_{\tilde{s}}(t)$};
\node[anchor=north west] at (quantizer_Q.-) {$v_{\bar{\tilde{s}}}(t)$};

\node[anchor=south west] at (out_I) {$v_{x_\ell}(t)$};
\node[anchor=north west] at (out_Q) {$v_{\bar{x}_\ell}(t)$};
% \draw (amp_I.out) to[open, v_=$v_{x_\ell}(t)$] (quantizer_I.+);

% \draw (quantizer_I.out) to[open, v=$v_{s_\ell}(t)$] ($(quantizer_I.+) + (0,0.785)$);

\end{circuitikz}
        }
        \caption{\label{fig:quadrature_impl}A single-ended op-amp implementation of a single quadrature stage from \Fig{fig:leapfrog-structure} together with a local quadrature control as in \Fig{fig:digital_control_zero_order_hold}.
        }
    \end{center}
\end{figure}
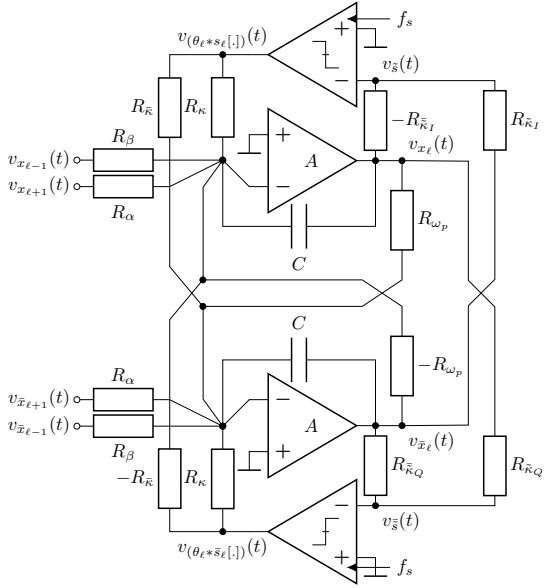
\subsection{Circuit Implementation Example}
To demonstrate that \Eq{eq:state_space_equation}-\Eq{eq:control_signal} can be implemented using conventional circuit techniques, \Fig{fig:quadrature_impl} shows a single-ended op-amp implementation of a single quadrature stage from \Fig{fig:leapfrog-structure}, with a local quadrature control as in \Fig{fig:digital_control_zero_order_hold}. 
The resistive and capacitive values follow from the $RC$ time constants as
$R_{\alpha}C = \alpha^{-1}$, $R_{\beta}C = \beta^{-1}$, $R_{\kappa}C = \kappa^{-1}$, $R_{\bar{\kappa}}C = \bar{\kappa}^{-1}$, 
$R_{\omega_n}C = \omega_n^{-1}$,  %from Kirchhoff's voltage and current laws follow
$R_{\tilde{\kappa}_I} R_I^{-1} = R_{\tilde{\kappa}_Q} R_Q^{-1} = \tilde{\kappa}$, and
$R_{\bar{\tilde{\kappa}}_I} R_I^{-1} = R_{\bar{\tilde{\kappa}}_Q} R_Q^{-1} = \bar{\tilde{\kappa}}$
for $R_I\eqdef R_{\tilde{\kappa}_I} - R_{\bar{\tilde{\kappa}}_I}$, $R_Q\eqdef R_{\tilde{\kappa}_Q} + R_{\bar{\tilde{\kappa}}_Q}$, and $R_{\tilde{\kappa}_I} \geq R_{\bar{\tilde{\kappa}}_I}$.
The presented circuit topology does have significant implementation challenges, in particular, the voltage dividers involving $(R_{\tilde{\kappa}_I}, R_{\bar{\tilde{\kappa}}_I}, R_{\tilde{\kappa}_Q}, R_{\bar{\tilde{\kappa}}_Q})$ could be replaced by multi-input comparators and the negative resistors may be managed in a differential setup. However, the purpose of \Fig{fig:quadrature_impl} is to demonstrate that quadrature \ac{CBADC} reduces to structures similar to \ac{CTSDM} circuit implementations.

\section{Simulation Results}\label{sec:simulation}

Behavioral simulations for multiple system specifications were done in Python using the cbadc toolbox \cite{cbadc:2022}.
Each simulation used a pair of full-scale input signals,
$u(t) = v_{\text{fs}} \cos(2 \pi f_s t)$ and $\bar{u}(t) = v_{\text{fs}} \sin(2 \pi f_s t)$ for $f_s = f_p - \omega_{\mathcal{B}} / (8 \pi)$, % $v_{\text{fs}}=?$,
$\phi_\kappa = \pi / 3$, $\tau_{\text{DC}} = 0$, and $v_{\text{fs}} = 1\text{V}$.
The final estimate of $u(t)$, in \Fig{fig:leapfrog-structure}, follows from \Eq{eq:estimate}.
The corresponding \ac{PSD}, after computing a $2^{14}$ point FFT, for $\text{\ac{OSR}}/N=(4/8,8/6)$ and multiple notch frequencies $f_n$ are shown in \Fig{fig:psd}. 
Note that, in contrast to the output of a \acp{CTSDM},
the \ac{CBADC}'s post-processing filter, \Eq{eq:estimate}, implicitly suppresses out-of-band frequencies. 
The \acp{SNR}, measured directly on the resulting \acp{PSD}, are approximately $83$ and $105$ dB, respectively, with variations within $\pm1$ dB between all notch frequencies. The reported \acp{SNR} also agree within $\pm1$ dB of the corresponding low-pass \ac{LF} system's performance. The low-pass \ac{PSD} is shown in blue in \Fig{fig:psd}.

A significant limiting factor for both \ac{BPSDM} and \ac{QSDM} is stability in the presence of component variations while targeting a wide tunable range.
The proposed \ac{QCBADC}'s sensitivity to component variations was tested by $256$ Monte Carlo simulations where each individual $(\alpha,\beta, \omega_n, \kappa_\phi,\bar{\kappa}_\phi,\tilde{\kappa}_\phi,\tilde{\bar{\kappa}}_\phi)$ value were drawn uniformly at random from within $\pm10$\% of their nominal values. For these simulations, the system was modeled using Verilog-A and simulated in Cadence Spectre with $\text{\ac{OSR}}=8, N=6$ and $f_n = f_s / 8$, which corresponded to $105$ dB nominal SNR, see \Fig{fig:psd} (bottom). The filter coefficients in \Eq{eq:estimate} were calculated from the actual system parametrization to avoid the additional filter mismatch error described in \Sec{sec:leapfrog}. 
The equivalent component variation scenario was also simulated for the low-pass \ac{LF} building block.
\begin{figure}
    \centering
    \vspace{0.125cm}
% \vspace{0.1cm}
\begin{tikzpicture}
    \pgfplotsset{
        grid=both, 
        % minor grid style = {densely dotted}, 
        % major grid style = {densely dotted}
    }
    \pgfplotsset{ylabel near ticks, xlabel near ticks}
    \pgfplotsset{ 
        legend cell align={left},
        legend style={
            at={(0.985,0.98)},
            anchor=north east,
            fill opacity=0.25,
            draw opacity=1,
            text opacity=1,
            nodes={scale=0.8, transform shape}
        }
    }
    % \begin{axis}[
    %     name=plot1,
    %     % xlabel={frequency [Hz]},
    %     % ylabel={PSD [dB]},
    %     yminorgrids,
    %     ymajorgrids,
    %     xminorgrids,
    %     xmajorgrids,
    %     xmin=0,
    %     xmax=0.5,
    %     ymin=-165,
    %     width=0.95\columnwidth,
    %     % height=\columnwidth/1.618,
    %     height=\columnwidth/2.6,
    %     xtick = {0, 0.0625, 0.125, 0.1875, 0.25, 0.3125, 0.375, 0.4375, 0.5},
    %     xticklabels = {$0$, ,$1/8$, ,$1/4$, ,$3/8$, , $1/2$},
    %     xticklabels = {},
    %     compat=1.16,
    %   cycle list={
    %     {black,mark=},
    %     {blue,mark=},
    %     {olive,mark=},
    %     {red,mark=}
    %     },
    %     % legend pos=north east
    %     ]
    %     \addplot+[mark=, color=black] table[col sep=comma, x=f_fp_0, y=fp_0] {./figures/psd_16_2.csv};
    %     \addplot+[mark=, color=blue] table[col sep=comma, x=f_fp_1, y=fp_1] {./figures/psd_16_2.csv};
        
    % \end{axis}
    \begin{axis}[
        name=plot2,
        % at=(plot1.below south), anchor=above north,
        % xlabel={frequency [Hz]},
        % ylabel={PSD [dB]},
        yminorgrids,
        ymajorgrids,
        xminorgrids,
        xmajorgrids,
        xmin=0,
        xmax=0.5,
        ymin=-205,
        width=0.95\columnwidth,
        % height=\columnwidth/1.618,
        height=\columnwidth/2.6,
        xtick = {0, 0.0625, 0.125, 0.1875, 0.25, 0.3125, 0.375, 0.4375, 0.5},
        xticklabels = {$0$, ,$1/8$, ,$1/4$, ,$3/8$, , $1/2$},
        xticklabels = {},
        compat=1.16, 
       cycle list={
        {black,mark=},
        {blue,mark=},
        {olive,mark=},
        {red,mark=}
        },
        % legend pos=north east
        ]
        
        % \addplot+[mark=, color=black] table[col sep=comma, x=f_fp_0, y=fp_0] {./figures/psd_4_4.csv};
        % \addplot+[mark=, color=red] table[col sep=comma, x=f_fp_1, y=fp_1] {./figures/psd_4_4.csv};
        % \addplot+[mark=, color=black] table[col sep=comma, x=f_fp_2, y=fp_2] {./figures/psd_4_4.csv};
        % \addplot+[mark=, color=red] table[col sep=comma, x=f_fp_3, y=fp_3] {./figures/psd_4_4.csv};
        
        \addplot+[mark=, color=black] table[col sep=comma, x=f_fp_0, y=fp_0] {./figures/psd_8_4.csv};
        \addplot+[mark=, color=red] table[col sep=comma, x=f_fp_1, y=fp_1] {./figures/psd_8_4.csv};
        \addplot+[mark=, color=black] table[col sep=comma, x=f_fp_2, y=fp_2] {./figures/psd_8_4.csv};
        \addplot+[mark=, color=red] table[col sep=comma, x=f_fp_3, y=fp_3] {./figures/psd_8_4.csv};
        
        \addplot+[mark=, color=blue] table[col sep=comma, x=f_bb, y=bb] {./figures/psd_8_4.csv};
        
    \end{axis}

    \node[rotate=90,xshift=-1.1cm, yshift=1.25cm] at (plot2.west) {$\text{V} / \sqrt{\text{Hz}}$ dB};

    \begin{axis}[
        name=plot3,
        at=(plot2.below south), anchor=above north,
        xlabel={$fT$},
        % cycle list name=auto,
        % xlabel={$fT$},
        % ylabel={PSD [dB]},
        yminorgrids,
        ymajorgrids,
        xminorgrids,
        xmajorgrids,
        xmin=0,
        xmax=0.5,
        ymin=-230,
        width=0.95\columnwidth,
        % height=\columnwidth/1.618,
        height=\columnwidth/2.6,
        xtick = {0, 0.0625, 0.125, 0.1875, 0.25, 0.3125, 0.375, 0.4375, 0.5},
        xticklabels = {$0$, ,$1/8$, ,$1/4$, ,$3/8$, , $1/2$},
        %xticklabels = {},
        % legend pos=north east
        ]
        
        \addplot[mark=, color=black] table[col sep=comma, x=f_fp_0, y=fp_0] {./figures/psd_6_8.csv};
        \addplot[mark=, color=olive] table[col sep=comma, x=f_fp_1, y=fp_1] {./figures/psd_6_8.csv};
        \addplot[mark=, color=black] table[col sep=comma, x=f_fp_2, y=fp_2] {./figures/psd_6_8.csv};
        \addplot[mark=, color=olive] table[col sep=comma, x=f_fp_3, y=fp_3] {./figures/psd_6_8.csv};
        \addplot[mark=, color=black] table[col sep=comma, x=f_fp_4, y=fp_4] {./figures/psd_6_8.csv};
        \addplot[mark=, color=olive] table[col sep=comma, x=f_fp_5, y=fp_5] {./figures/psd_6_8.csv};
        \addplot[mark=, color=black] table[col sep=comma, x=f_fp_6, y=fp_6] {./figures/psd_6_8.csv};
        \addplot[mark=, color=olive] table[col sep=comma, x=f_fp_7, y=fp_7] {./figures/psd_6_8.csv};
        
        \addplot[mark=,color=blue] table[col sep=comma, x=f_bb, y=bb] {./figures/psd_6_8.csv};

        % \addlegendentry{calibrated}
    \end{axis}
\end{tikzpicture}
\vspace{-0.5cm}
    \caption{\label{fig:psd} 
    \ac{PSD} for different notch frequencies. The black, red, and yellow lines correspond to \acp{PSD} of \Eq{eq:estimate} from \acp{QCBADC} designed for $\text{\ac{OSR}}/N=(4/8,8/6)$ (top, bottom) and positioned at different notch filter frequencies $f_n$. 
    Similarly, the blue lines are the \acp{PSD} of a corresponding low-pass \ac{CBADC} building block. 
    }
\end{figure}
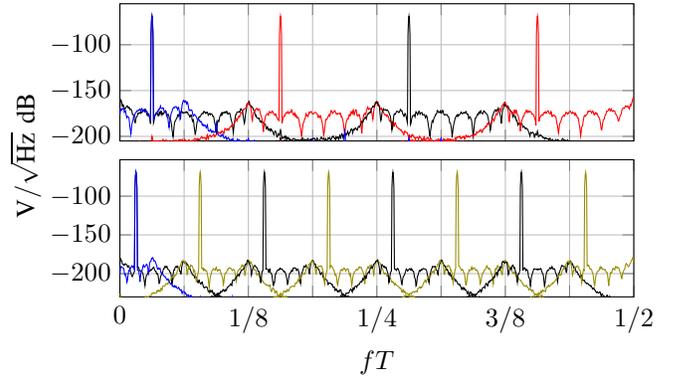
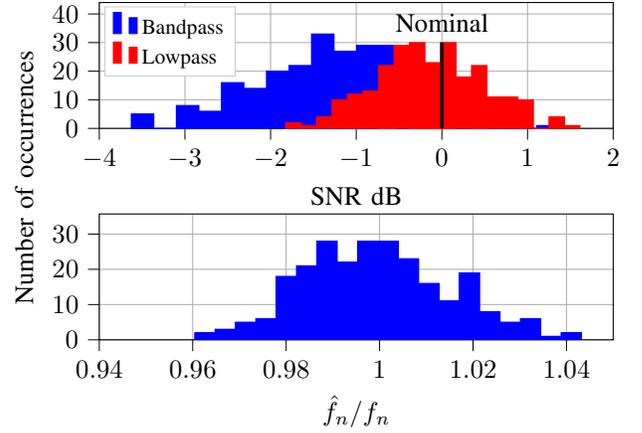
\begin{figure}
    \centering
    % \vspace{0.05cm}
\vspace{-0.15cm}
\begin{tikzpicture}
    \pgfplotsset{
        grid=both, 
        % minor grid style = {densely dotted}, 
        % major grid style = {densely dotted}
    }
    \pgfplotsset{ylabel near ticks, xlabel near ticks}
    \pgfplotsset{ 
        legend cell align={left},
        legend style={
            at={(0.985,0.98)},
            anchor=north east,
            fill opacity=0.25,
            draw opacity=1,
            text opacity=1,
            nodes={scale=0.8, transform shape}
        }
    }

\definecolor{darkgray176}{RGB}{176,176,176}
\definecolor{lightgray204}{RGB}{204,204,204}

\begin{axis}[
name=plot4,
at=(plot3.below south), anchor=above north,
width=0.95\columnwidth,
height=\columnwidth/2.72,
legend cell align={left},
legend style={
  fill opacity=0.8,
  draw opacity=1,
  text opacity=1,
  at={(0.01,0.97)},
  anchor=north west,
  draw=lightgray204
},
tick align=outside,
tick pos=left,
x grid style={darkgray176},
xlabel={SNR dB},
% ylabel=No. occurrences,
xmin=-4, xmax=2,
xtick style={color=black},
y grid style={darkgray176},
ymin=0, ymax=44,
ytick style={color=black}
]

% BANDPASS
\draw[draw=blue,fill=blue] (axis cs:-3.62619059305403,0) rectangle (axis cs:-3.36319891774028,5);
\addlegendimage{ybar,ybar legend,draw=blue,fill=blue}
\addlegendentry{Bandpass}
\draw[draw=blue,fill=blue] (axis cs:-3.36319891774028,0) rectangle (axis cs:-3.10020724242652,0);
\draw[draw=blue,fill=blue] (axis cs:-3.10020724242652,0) rectangle (axis cs:-2.83721556711277,8);
\draw[draw=blue,fill=blue] (axis cs:-2.83721556711277,0) rectangle (axis cs:-2.57422389179901,6);
\draw[draw=blue,fill=blue] (axis cs:-2.57422389179901,0) rectangle (axis cs:-2.31123221648526,16);
\draw[draw=blue,fill=blue] (axis cs:-2.31123221648526,0) rectangle (axis cs:-2.0482405411715,14);
\draw[draw=blue,fill=blue] (axis cs:-2.0482405411715,0) rectangle (axis cs:-1.78524886585774,20);
\draw[draw=blue,fill=blue] (axis cs:-1.78524886585774,0) rectangle (axis cs:-1.52225719054399,22);
\draw[draw=blue,fill=blue] (axis cs:-1.52225719054399,0) rectangle (axis cs:-1.25926551523023,33);
\draw[draw=blue,fill=blue] (axis cs:-1.25926551523023,0) rectangle (axis cs:-0.996273839916477,27);
\draw[draw=blue,fill=blue] (axis cs:-0.996273839916477,0) rectangle (axis cs:-0.733282164602722,29);
\draw[draw=blue,fill=blue] (axis cs:-0.733282164602722,0) rectangle (axis cs:-0.470290489288966,29);
\draw[draw=blue,fill=blue] (axis cs:-0.470290489288966,0) rectangle (axis cs:-0.20729881397521,15);
\draw[draw=blue,fill=blue] (axis cs:-0.20729881397521,0) rectangle (axis cs:0.0556928613385459,12);
\draw[draw=blue,fill=blue] (axis cs:0.0556928613385459,0) rectangle (axis cs:0.318684536652301,13);
\draw[draw=blue,fill=blue] (axis cs:0.318684536652301,0) rectangle (axis cs:0.581676211966057,2);
\draw[draw=blue,fill=blue] (axis cs:0.581676211966057,0) rectangle (axis cs:0.844667887279813,2);
\draw[draw=blue,fill=blue] (axis cs:0.844667887279813,0) rectangle (axis cs:1.10765956259357,0);
\draw[draw=blue,fill=blue] (axis cs:1.10765956259357,0) rectangle (axis cs:1.37065123790732,1);

% BASEBAND
\draw[draw=red, opacity=0.8,fill=red] (axis cs:-1.82533651911319,0) rectangle (axis cs:-1.64443147197269,2);
\addlegendimage{ybar,ybar legend,draw=red,fill=red}
\addlegendentry{Lowpass}
\draw[draw=red, opacity=0.8,fill=red] (axis cs:-1.64443147197269,0) rectangle (axis cs:-1.46352642483219,1);
\draw[draw=red, opacity=0.8,fill=red] (axis cs:-1.46352642483219,0) rectangle (axis cs:-1.28262137769168,4);
\draw[draw=red, opacity=0.8,fill=red] (axis cs:-1.28262137769168,0) rectangle (axis cs:-1.10171633055118,10);
\draw[draw=red, opacity=0.8,fill=red] (axis cs:-1.10171633055118,0) rectangle (axis cs:-0.920811283410679,12);
\draw[draw=red, opacity=0.8,fill=red] (axis cs:-0.920811283410679,0) rectangle (axis cs:-0.739906236270176,13);
\draw[draw=red, opacity=0.8,fill=red] (axis cs:-0.739906236270176,0) rectangle (axis cs:-0.559001189129674,22);
\draw[draw=red, opacity=0.8,fill=red] (axis cs:-0.559001189129674,0) rectangle (axis cs:-0.378096141989172,29);
\draw[draw=red, opacity=0.8,fill=red] (axis cs:-0.378096141989172,0) rectangle (axis cs:-0.197191094848669,30);
\draw[draw=red, opacity=0.8,fill=red] (axis cs:-0.197191094848669,0) rectangle (axis cs:-0.0162860477081672,23);
\draw[draw=red, opacity=0.8,fill=red] (axis cs:-0.0162860477081672,0) rectangle (axis cs:0.164618999432335,30);
\draw[draw=red, opacity=0.8,fill=red] (axis cs:0.164618999432335,0) rectangle (axis cs:0.345524046572837,18);
\draw[draw=red, opacity=0.8,fill=red] (axis cs:0.345524046572837,0) rectangle (axis cs:0.52642909371334,22);
\draw[draw=red, opacity=0.8,fill=red] (axis cs:0.52642909371334,0) rectangle (axis cs:0.707334140853842,11);
\draw[draw=red, opacity=0.8,fill=red] (axis cs:0.707334140853842,0) rectangle (axis cs:0.888239187994344,11);
\draw[draw=red, opacity=0.8,fill=red] (axis cs:0.888239187994344,0) rectangle (axis cs:1.06914423513485,10);
\draw[draw=red, opacity=0.8,fill=red] (axis cs:1.06914423513485,0) rectangle (axis cs:1.25004928227535,0);
\draw[draw=red, opacity=0.8,fill=red] (axis cs:1.25004928227535,0) rectangle (axis cs:1.43095432941585,4);
\draw[draw=red, opacity=0.8,fill=red] (axis cs:1.43095432941585,0) rectangle (axis cs:1.61185937655635,1);

\path [draw=black, very thick]
(axis cs:0,0)
--(axis cs:0,30) node[above]{Nominal};

\end{axis}

\node[rotate=90,xshift=-1.5cm, yshift=1cm] at (plot4.west) {Number of occurrences};

\begin{axis}[
name=plot5,
at=(plot4.below south), anchor=above north,
width=0.95\columnwidth,
height=\columnwidth/2.72,
legend cell align={left},
legend style={
  fill opacity=0.8,
  draw opacity=1,
  text opacity=1,
  at={(0.03,0.97)},
  anchor=north west,
  draw=lightgray204
},
tick align=outside,
tick pos=left,
x grid style={darkgray176},
xlabel={$\hat{f}_n / f_n$},
xmin=0.94, xmax=1.05,
xtick style={color=black},
y grid style={darkgray176},
ymin=0, ymax=35.7,
ytick style={color=black}
]
\draw[draw=blue,fill=blue] (axis cs:0.960433779411535,0) rectangle (axis cs:0.964790261834659,2);
\draw[draw=blue,fill=blue] (axis cs:0.964790261834659,0) rectangle (axis cs:0.969146744257783,3);
\draw[draw=blue,fill=blue] (axis cs:0.969146744257783,0) rectangle (axis cs:0.973503226680906,5);
\draw[draw=blue,fill=blue] (axis cs:0.973503226680906,0) rectangle (axis cs:0.97785970910403,6);
\draw[draw=blue,fill=blue] (axis cs:0.97785970910403,0) rectangle (axis cs:0.982216191527154,18);
\draw[draw=blue,fill=blue] (axis cs:0.982216191527154,0) rectangle (axis cs:0.986572673950278,21);
\draw[draw=blue,fill=blue] (axis cs:0.986572673950278,0) rectangle (axis cs:0.990929156373402,28);
\draw[draw=blue,fill=blue] (axis cs:0.990929156373402,0) rectangle (axis cs:0.995285638796525,22);
\draw[draw=blue,fill=blue] (axis cs:0.995285638796525,0) rectangle (axis cs:0.999642121219649,28);
\draw[draw=blue,fill=blue] (axis cs:0.999642121219649,0) rectangle (axis cs:1.00399860364277,28);
\draw[draw=blue,fill=blue] (axis cs:1.00399860364277,0) rectangle (axis cs:1.0083550860659,23);
\draw[draw=blue,fill=blue] (axis cs:1.0083550860659,0) rectangle (axis cs:1.01271156848902,16);
\draw[draw=blue,fill=blue] (axis cs:1.01271156848902,0) rectangle (axis cs:1.01706805091214,11);
\draw[draw=blue,fill=blue] (axis cs:1.01706805091214,0) rectangle (axis cs:1.02142453333527,19);
\draw[draw=blue,fill=blue] (axis cs:1.02142453333527,0) rectangle (axis cs:1.02578101575839,8);
\draw[draw=blue,fill=blue] (axis cs:1.02578101575839,0) rectangle (axis cs:1.03013749818152,5);
\draw[draw=blue,fill=blue] (axis cs:1.03013749818152,0) rectangle (axis cs:1.03449398060464,6);
\draw[draw=blue,fill=blue] (axis cs:1.03449398060464,0) rectangle (axis cs:1.03885046302776,1);
\draw[draw=blue,fill=blue] (axis cs:1.03885046302776,0) rectangle (axis cs:1.04320694545089,2);
\end{axis}

\end{tikzpicture}
    \caption{\label{fig:hist} Histograms showing variation in \ac{SNR} relative to the nominal performance of $105$ dB (top) and estimated $\hat{f}_n$ normalized to nominal $f_n=f_s / 8$ (bottom) after 256 Monte Carlo simulations with up to $\pm 10\%$ variations in each $(\alpha,\beta,\omega_n,\kappa_\phi,\bar{\kappa}_\phi,\tilde{\kappa}_\phi,\tilde{\bar{\kappa}}_\phi)$ parameter.}
\end{figure}
For the low-pass \ac{LF}, the component variations resulted in $3$ realizations being unstable and, therefore, significantly reduced \ac{SNR}. In contrast, none of the \ac{QCBADC} realizations resulted in instability. 
Histograms of the resulting \ac{SNR} performance and estimated notch frequencies $\hat{f}_n$, excluding the unstable low-pass realizations, are given in \Fig{fig:hist}. The results show \acp{SNR} and notch frequencies $\hat{f}_n$ in the ranges of $(-4,2)$~dB, and $\pm 5$\% from their respective nominal values.  
We conjecture, that
the average reduction of $1$~dB \ac{SNR}, in the \ac{QCBADC} case, originates from the fact that the quadrature stages in \Fig{fig:leapfrog-structure} may lose amplification, in the transfer function from $(u(t), \bar{u}(t))$ to $(x_N(t), \bar{x}_N(t))$, by the combined effect of changes in bandwidth and non-aligning resonance frequencies per stage. This differs from the low-pass \ac{LF} case where the same component variation simulation shows no significant change to average \ac{SNR} as effectively only the bandwidth per stage varies.
How to calibrate the \ac{DE}, as was effectively assumed in the above simulations, is further described in \cite{MMBFL:22}. 

\section{Conclusions}

The \ac{QCBADC} design principle extends any low-pass \acp{CBADC} into a quadrature version centered around a desired notch frequency without loss of \ac{SNR} or stability margin. Nominal performance and tuning range are verified by behavioral simulations of different system orders and \acp{OSR}. 
Robustness is demonstrated by electrical Monte Carlo simulations, with up to $10$\% component variations, resulting in a performance range of approximately $(-4, 2)$ dB, with respect to the nominal \ac{SNR}. These results demonstrate that quadrature \ac{CBADC} is a good alternative for the implementation of RF digitizers with continuous tuning of the notch frequency, opening the doors to more efficient realization of software-defined radios.

\newpage

\newcommand{\norcas}{IEEE Nordic Circuits and Syst. Conf. (NorCAS)}


\begin{thebibliography}{00}
    % Journals and conference names
    \newcommand{\tcasi}{IEEE Trans. Circuits Syst. I, Reg. Papers}
    \newcommand{\tcasiiadsp}{IEEE Trans. Circuits Syst. II: Analog Digit. Signal Process.}
    \newcommand{\jssc}{IEEE J. Solid-State Circuits}
    \newcommand{\iscas}{Proc. IEEE Int. Symp. Circuits Syst. (ISCAS)}
    \newcommand{\jestcs}{IEEE J. Emerg. Sel. Topics Circuits Syst.}
    % Citation commands
    \newcommand{\phd}[3]{#1, \enquote{#2,} Ph.D. dissertation, #3.}
    % \bibitem{LW:15} H.-A. Loeliger and G. Wilckens, \enquote{Control-based analog-to-digital conversion without sampling and quantization,} in \textit{2015 Information Theory \& Applications Workshop (ITA),}, San Diego, CA, pp. 119-122.
        %% JMdelaRosa
\bibitem{saye20}
A.~Sayed {\em et~al.}, ``{A 1.5-to-3.0GHz Tunable RF Sigma-Delta ADC With a
  Fixed Set of Coefficients and a Programmable Loop Delay},'' {\em IEEE
  Transactions on Circuits and Systems - II: Express Briefs}, vol.~67,
  pp.~1559--1563, September 2020.

\bibitem{ghae21}
H.~Ghaedrahmati, J.~Zhou, and R.~B. Staszewski, ``{A 38.6-fJ/Conv.-Step
  Inverter-Based Continuous-Time Bandpass $\Delta\Sigma$ ADC in 28 nm Using
  Asynchronous SAR Quantizer},'' in {\em IEEE Transactions on Circuits and Systems
  - II: Express Briefs}, vol.~68, pp.~3113--3117, September 2021.

\bibitem{jie21}
L.~Jie {\em et~al.}, ``{A 100MHz-BW 68dB-SNDR Tuning-Free Hybrid-Loop DSM with
  an Interleaved Bandpass Noise-Shaping SAR Quantizer},'' {\em IEEE ISSCC
  Digest of Technical Papers}, February 2021.

\bibitem{moli14}
G.~Molina {\em et~al.}, ``{LC-Based Bandpass Continuous-Time Sigma-Delta
  Modulators with Widely Tunable Notch Frequency},'' {\em IEEE Trans. on
  Circuits and Systems -- I: Regular Papers}, pp.~1442--1455, May 2014.
  %% JMdelaRosa

  
% \bibitem{KWJWH:08}
% S.~-B. Kim, T.~D. Werth, S.~Joeres, R.~Wunderlich and S.~Heinen, \enquote{Effect of Mismatched Loop Delay in Continuous-Time Complex Sigma-Delta Modulators,} in {\em IEEE Transactions on Circuits and Systems
%   - II: Express Briefs}, vol. 55, no. 10, pp. 996-1000, Oct. 2008.


\bibitem{M:20}{H. Malmberg, \enquote{Control-Bounded Converters,} Ph.D. dissertation no. 27025, ETH Zürich, 2020}

\bibitem{MWL:21}{H. Malmberg, G. Wilckens, and H.-A. Loeliger, \enquote{Control-bounded analog-to-digital conversion,} \textit{Circuits, Syst. Signal Process.,}  vol. 41, no. 3, pp. 1223-1254, Mar. 2022.}

\bibitem{ST:05}
R. Schreier, G. C. Temes, \enquote{Bandpass and Quadrature Delta‐Sigma Modulation,} in \textit{Understanding Delta-Sigma Data Converters}. New York, NY USA: Wiley, 2005, pp. 172-177.

% \bibitem{MWL:21} H. Malmberg, G. Wilckens, and H.-A. Loeliger, \enquote{Control-bounded analog-to-digital conversion,} \textit{Circuits, Syst. Signal Process.,}  vol. 41, no. 3, pp. 1223-1254, Mar. 2022.
% \bibitem{Widrow:75} B. Widrow, J.R. Glover, J.M. McCool, J. Kaunitz, C.S. Williams, R.H. Hearn, J.R. Zeidler, Jr. Eugene Dong, and R.C. Goodlin, \enquote{Adaptive noise cancelling: principles and applications,} \textit{Proceedings of the IEEE,} vol. 63, no. 12, pp. 1692-1716, 1975.

\bibitem{MMBFL:22} H. Malmberg, T. Mettler, T. Burger, F. Feyling, H.-A. Loeliger, \enquote{Calibrating control-bounded ADCs,} arXiv:2211.06741v1.


% \bibitem{FMWLY:2022}{F. Feyling, H. Malmberg, C. Wulff, H.-A Loeliger, and T. Ytterdal, \enquote{High-level comparison of control-bounded A/D converters and continuous-time sigma-delta modulators,} in \textit{Nordic Circuits and Systems Conference (NORCAS)}, Oslo, pp. xx, Oct. 2022.}

\bibitem{FMWLY:2022} F. Feyling, H. Malmberg, C. Wulff, H.-A. Loeliger, and T. Ytterdal, \enquote{High-level comparison of control-bounded A/D converters and continuous-time sigma-delta modulators,} in \textit{\norcas}, Oslo, Norway, Oct. 2022, pp. 1-5.

% \bibitem{FMWLY:2022}{----------, \enquote{High-level comparison of control-bounded A/D converters and continuous-time sigma-delta modulators,} in \textit{2022 Nordic Circuits and Systems Conference (NORCAS)}, Oslo, Oct. 2022, unpublished.}

% \bibitem{LDHKPK:2007} H.-A. Loeliger, J. Dauwels, Junli Hu, S. Korl, Li Ping, and F. R. Kschischang, \endquote{The factor graph approach to model-based signal processing,} \textit{Proceedings of the IEEE,} vol. 95, no. 6, pp. 1295-1322, June 2007.
% \bibitem{Haykin:2002} S. Haykin, Adaptive filter theory, 4th ed. Upper Saddle River, NJ: Prentice Hall, 2002.
%\bibitem{MMDR:14} G. Molina-Salgado, A. Morgado, G. J. Dolecek, J. M. de la Rosa, \enquote{LC-Based Bandpass Continuous-Time Sigma-Delta Modulators With Widely Tunable Notch Frequency,} \textit{\tcasi}, vol. 61, no. 5, pp. 1442-1455, May. 2014.

\bibitem{cbadc:2022}{H. Malmberg, Control-Bounded A/D Conversion Toolbox (cbadc), https://github.com/hammal/cbadc.git, v0.2.2, 2022.}
\end{thebibliography}
\end{document}